# Study of the structural defects on carbon nanotubes in metal matrix composites processed by severe plastic deformation

Katherine Aristizabal[1], Andreas Katzensteiner[2], Andrea Bachmaier[2], Frank Mücklich[1], and Sebastian Suarez[1,*].

[1]Chair of Functional Materials, Department of Materials Science, Saarland University. D-66123, Saarbrücken, Germany.
[2]Erich Schmid Institute of Materials Science, Austrian Academy of Sciences, Jahnstrasse 12, A-8700 Leoben, Austria.

## Abstract

Carbon nanotubes (CNT) have been recently proposed as stabilizers against grain growth that can happen even at low temperature inputs in nano-crystalline and ultrafine-grained materials obtained by severe plastic deformation. In this study, we analyzed the evolution of the structural defects on the nanotubes in CNT-reinforced nickel matrix composites with different reinforcement weight fractions. The composites were processed by high pressure torsion, and we used Raman spectroscopy as the main characterization technique. The results indicate that for CNT subjected to highly energetic processing, it is not sufficient to analyze only the $I_D/I_G$ ratio (as proposed in the available literature), but it is also necessary to evaluate the shifting of the G band, which traces the amorphization trajectory undergone by the CNT. Furthermore, we observed that the deformation suffered by the CNT is related to the accumulated strain and varies with the partial CNT fractions of these composites. This is related to their capacity to withstand the plastic strain that occurs during deformation. In addition, the defective state reaches a saturation before achieving the saturation in the microstructural refinement. These results will help to efficiently optimize the processing of this type of engineering composites.

*Corresponding author. Tel: +49 (681) 30270538. E-mail: s.suarez@mx.uni-saarland.de (Sebastian Suarez)

## 1. Introduction

Carbon nanotubes (CNT) possess superior characteristics (outstanding physical properties, low weight and high aspect ratio), which render them as suitable candidates to be used as reinforcement in composite materials. In the past years, there have been several reports where they are used as reinforcements in metal matrix composites (MMC) with different metallic matrices, aiming mainly to improve the electrical and mechanical properties. An improvement in the mechanical properties has been observed by adding different volume fractions of CNT to different metallic matrices (e.g. Al [1–8], Cu [9–11] and Ni [12–14]). Nevertheless, the processing route also dictates to what extent the properties are improved, where the agglomerates and a weak interface can lead to a lower improvement [15] or even to a deterioration thereof [16]. CNT have been also used in MMC as stabilizing phases against grain growth in materials subjected to severe plastic deformation (SPD)[17,18]. This was based on their ability to pin the grain boundaries in coarse-grained materials under thermal inputs, enabling the control of the final microstructure [19] and subsequently, the tailoring of their mechanical properties [12]. Particularly, the boundary drag effect becomes more evident in composites manufactured by solid-state routes (e.g. powder metallurgy), since the reinforcement can only be placed on grain boundaries, as opposed to other chemical synthesis routes (e.g. molecular-level mixing). A thorough literature review about the use of CNT in MMCs is beyond the scope of this manuscript and can be found in [20] and the references therein.

As previously demonstrated [21], the particle distribution homogenization in MMCs is achievable using high pressure torsion (HPT), which is essential in the case of CNT, given their predisposition to form agglomerates due to van der Waals interactions. The homogeneous distribution of the particles within the matrix in MMCs is very important to avoid anisotropic behavior during mechanical loading. Additionally, in the case of CNT-reinforced composites, it is also of paramount importance that the CNT keep their microstructural features as unaltered as possible after processing. This would ensure to some extent that the CNT will retain their physical properties and could avoid chemical interactions with the metallic matrix. These chemical interactions are detrimental to the mechanical performance, since they generate brittle phases (i.e. carbides) which stem from the degradation of the C-containing phase. Even though some authors correctly state that the presence of carbides enhance the interface in composites, this would be a detrimental feature in two ways. First, the aforementioned brittle interface might crack after applying a certain stress, thus rendering the load transfer negligible. Second, by degrading the CNT, their intrinsic physical properties are severely compromised. Therefore, a processing route in which the structural state of the CNT is retained and a seamless interface with the matrix is achieved becomes extremely important.

Different kind of defects (e.g. point defects or dangling bonds) can be present in CNT after highly energetic processes such as ball milling [22–24], which can lead to interaction between the CNT and the matrix. In the same way, seeking the refinement of the microstructure by means of HPT, the samples can be subjected simultaneously to high pressures and high strains, and although the CNT are surrounded by the softer matrix and can –to some extent– be protected by it, they might be degraded during the processing. It is widely recognized that Raman spectroscopy is a very powerful and versatile technique for the characterization of carbon-based materials. By analyzing the change of certain characteristic bands of $sp^2$ carbons (D, G and G' bands and their intensity ratios) and certain descriptive features (peak central position and full width at half maximum – FWHM, $\Gamma$ – of the G band), it is possible to obtain a fairly accurate picture of the structural state of the carbonaceous phase. To the best of our understanding, there are no thorough studies of the damage suffered by CNT after SPD using Raman Spectroscopy. Though certain reports present Raman spectra of deformed composites, the analysis is extended (in the best cases) to an observation of the so-called defect index ($I_D/I_G$) [17,25,26], which would not present a full overview of the structural state. Specifically, Tokunaga and coworkers [25] studied the HPT deformation of CNT-reinforced Al composites at 2.5 GPa and 30 turns. They report that there is an increased D-band intensity due to structural damage generated by bending and/or breaking of the CNT. However, by analyzing the presented Raman spectra, an increase in the G-band width and a strong upshifting is clearly noticeable. This might indicate a possible amorphization of the graphitic structure. This was also observed in [26], even after the pre-processing, but was not discussed. In this case, the authors only focus on a brief analysis of the $I_D/I_G$ ratio. Thus, in light of the available literature, a thorough study of the influence of HPT on the CNT structure becomes very important.

In the present work, we obtained CNT-reinforced nickel matrix composites via HPT, and evaluated the influence of the processing parameters on the nanotubes structural state for different CNT weight fractions. These composites were sintered by means of hot uniaxial pressing (HUP) and afterwards subjected to HPT at room temperature using different number of turns (T) applying a pressure of 4 GPa.

## 2. Experimental

### 2.1. Manufacturing and HPT processing of CNT/Ni composites

Different CNT weight fractions were analyzed, namely: 0.5 wt.% (2.4 vol.%), 1 wt.% (4.7 vol.%), and 2 wt.% (9 vol.%), the latter being the maximum reasonable amount of CNT to be used in CNT/Ni composites as determined in previous studies [12]. The manufacturing of the composites started with the dispersion of multi-wall carbon nanotubes (MWCNT) (CCVD grown, Graphene Supermarket, USA. Density 1.84 g/cm$^3$, diameter: 50-85 nm, length: 10-15 µm, carbon purity: >94%) in ethylene glycol EG (MWCNT/EG concentration ratio at 0.2 mg/ml). Afterwards, a mixture with nickel dendritic powder (Alfa Aesar, Mesh 325 (45 µm), density 8.91 g/cm$^3$) was produced by means of a homogenizer (WiseTis, Witeg) to disperse the larger agglomerates, and an ultrasonic bath (Sonorex Super RK 514 BH, Bandelin, 860 W, 35 kHz) to disperse the smaller ones. The dispersion was performed following the process described in [27]. The solvent was evaporated in a ventilated furnace at 150 °C and the powder was carefully grinded using an agate mortar to obtain more uniformly distributed particles and therefore to ease the compaction of the green pellets (990 MPa). After the densification process at 750 °C in vacuum (2 x 10$^{-6}$ mbar) for 2.5 h, using a hot uniaxial press (264 MPa), the resulting samples with different compositions were subjected to HPT at room temperature using different number of turns (T = 1, 4, 10 and 20) applying 4 GPa of pressure. An additional set of samples with 1wt.% and 2wt.% were processed in HPT using 1 rotation, 4 GPa and 7 GPa, in order to evaluate the effect of pressure on the damage. Raman measurements were performed in the center of the samples, where no deformation caused by strain should take place, in order to analyze the effect of the pressure subjected to the MWCNT.

The samples were cut in halves and embedded using conductive resin. All the analyses were carried out along the radial direction (Figure 1). The shear strain in HPT is calculated by:

$$\gamma = \frac{2\pi T}{t} r \quad \text{Eq. 1}$$

Where T is the number of revolutions, r is the distance from the center of the rotation axis and t is the thickness of the sample. If one assumes a von Mises yield locus [28], the equivalent strain applied in HPT can be calculated by:

$$\varepsilon_{vM} = \frac{2\pi T}{\sqrt{3}\, t} r \qquad \text{Eq. 2}$$

The sample diameter was 8 mm in all cases and the thickness varied from 0.6 mm to 1 mm, so the equivalent strain was specifically calculated for each sample radius.

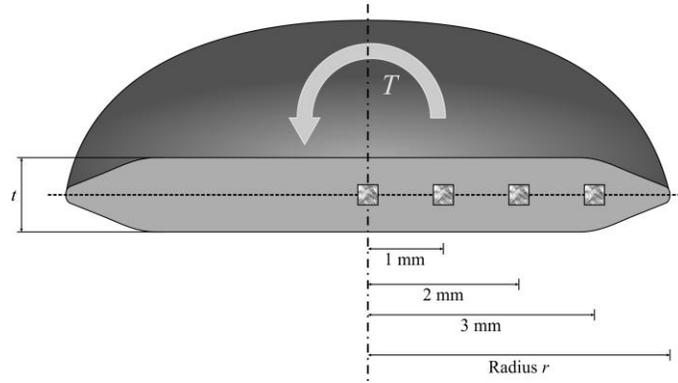

*Figure 1 – Schematic view of the measurement positions in the cross section of the composites.*

2.2. Structural Characterization of the CNT

Raman spectra were acquired with an InVia™ Raman spectrometer (Renishaw) using a 2400 lines/mm grating and an Nd:YAG laser with an excitation wavelength of 532 nm (2.33 eV) through a 50x objective (numerical aperture of 0.9) with a spectral resolution of 1.2 cm$^{-1}$. Each spectrum is the average of 3 measurements performed on the same spot (spot size of 3 µm). Measurements along the radii were carried out every 1 mm in order to visualize the evolution of the data along the middle plane and the results were compared to those of the reference samples in the "as sintered" condition. Following the analysis proposed by N. Souza et al. [29], it was confirmed that no laser-induced modification of the CNT occurs up to a laser power of 0.5 mW. According to this, the laser power was set for all the measurements to 0.2 mW. Information about the height of the G, D and G' bands were taken directly from the normalized data and their ratios were computed and plotted against the corresponding equivalent strain values. A Lorentzian fit of the spectra was used in order to get information about position of the G-band as well as its full width at half maximum. The analysis was carried out by studying the evolution of different descriptive parameters such as the full width at half maximum of the G band $\Gamma_G$ (crystallinity-related), $I_D/I_G$ ratio (defect-related), and the G band position (amorphization-stage related). In all the cases, the data was recorded along the middle plane in order to avoid the effect of the microstructural gradients along the height of the specimens on the results, as shown in Figure 1 [30].

## 3. Results and discussion

The microstructural refinement of these composites was studied and reported elsewhere (Katzensteiner A. et al. IOP Conf. Series: Materials Science and Engineering 194 (2017) 012019 doi:10.1088/1757-899X/194/1/012019). During HPT a pronounced microstructural refinement takes place, which implies that the CNT present in the composites also undergo strong deformations during processing.

All the results indicate that the larger variations of the Raman parameters occur during the early stages of processing, where the application of high compressive loads takes place. This can be observed in all the cases, when comparing the results to those of the reference samples.

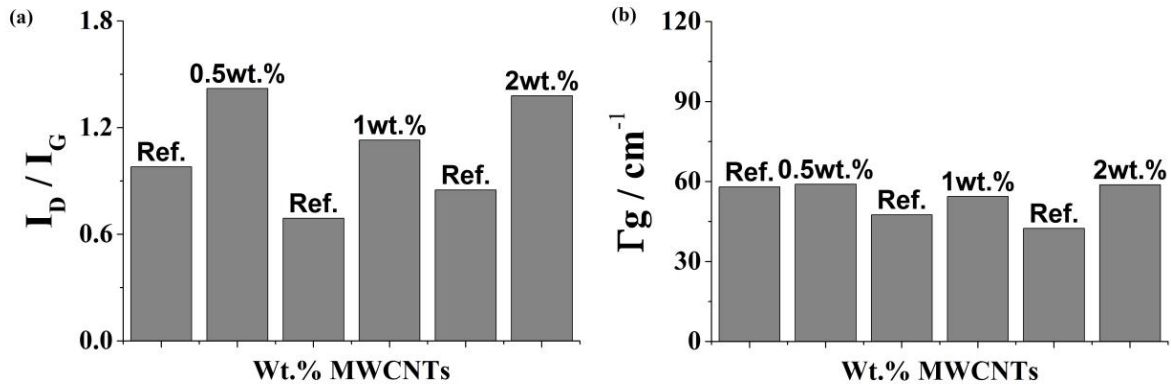

*Figure 2 - Effect of the applied Pressure during HPT (taken at the center of the sample, $\varepsilon_{vM} = 0$) after 1 turn. (a) Defect index $I_D/I_G$ and, (b) full width at half maximum (FWHM) of the G band.*

Figure 2 shows some qualitative markers of the CNT structural state that are influenced mainly by the applied pressure during HPT. According to Equation 2, the measurements taken at the center of the sample (r = 0) would not present accumulated von Mises strain ($\varepsilon_{vM} = 0$). By studying the case for 1 wt.% and 2 wt.% samples, it becomes evident that the application of pressure increases the structural disorder in the CNT (Figure 2a). Additionally, the peak width is also increased, indicating a reduction in the crystallinity (Figure 2b). The broadening of the G band indicates a reduction in the resonance-induced vibration extinction time [31], meaning that there is either a rise in the amount of structural defects, a shortening of the CNT or a combination thereof.

*Table 1 – Mean CNT agglomerate sizes in the as-sintered (initial) and post-deformation (final) state for a subset of samples measured at the center of the sample.*

| Sample | Initial agglomerate size /µm | Final agglomerate size /µm |
| --- | --- | --- |
| 1 wt.%_1T_4GPa | 4.1 ± 1.5 | 0.8 ± 0.3 |
| 1 wt.%_1T_7GPa |  | 0.4 ± 0.2 |
| 2 wt.%_1T_4GPa | 6.6 ± 2.4 | 0.4 ± 0.2 |
| 2 wt.%_1T_7GPa |  | 0.3 ± 0.1 |

Concerning the CNT agglomerate size, a significant reduction in size is clearly noticeable as shown in Table 1. The change in the mean agglomerate size shows a drop of one order of magnitude after only one turn in HPT. The values stay in the sub-micron range, which might be more useful when a dispersion-strengthening effect is needed, as opposed to the as-sintered state with µm-ranged clusters. As an example (Figure 3), the deformation induced by a 20 turns HPT processing rendered a more homogeneous and finer CNT cluster distribution throughout the metallic matrix. This has been already reported for hard, ceramic particles in soft metallic matrices [21].

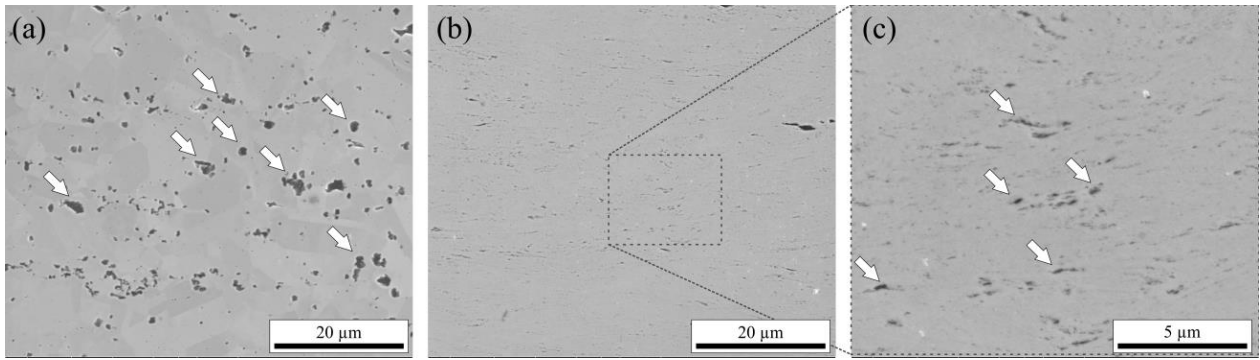

*Figure 3 – Electron micrographs of a 1 wt.% CNT composite (a) at the undeformed initial state and (b) after 20 turns of deformation, taken at 3 mm from the center. (c) Magnification of a region of interest. As a guide, the white*

*arrows indicate some regions with CNT agglomerates (This type of composites encloses the CNT in the cavities as observed by EDX, not shown here). The reduction of the mean agglomerate size is clearly noticeable.*

When applying high compressive loads, a relative gliding between adjacent CNT within the agglomerates takes place. This derives in a significant reduction in the cluster size and a better detachment of the impurities when applying higher compressive loads (7 GPa). This is confirmed by these results, where the 7 GPa processed samples present a smaller mean agglomerate size than those processed with 4 GPa. This feature adds a significantly practical solution to a prevalent complication in the manufacturing of CNT-reinforced metal matrix composites, where agglomeration is one of the major concerns towards practical applicability.

*Effect of the accumulated strain (at 4GPa)*

The study of the effect of accumulated strain on the CNT has been approached by analysing three different indicators ($\Gamma_G$, $I_D/I_G$ ratio and G band position) within an equivalent strain range of $\varepsilon_{vM} = 0 - 375$.

The crystallinity-related index $\Gamma_G$ shows a constant increase in its value in all cases (Figure 4a), indicating an increase in amorphization of the nanotubes. Interestingly, all samples show a saturation in the broadening at approximately 90 cm$^{-1}$ at high deformation grades. This indicates a greater structural disorder and a likely shortening of the CNT, as explained before. Furthermore, it is evident that the broadening happens during the early stages of deformation ($\varepsilon_{vM} < 5$). This is related to the fact that, as the agglomerates are detached and new smaller agglomerates appear, the load is distributed among a larger number of clusters. Thus, the effective pressure applied on each of the agglomerates would be higher when a higher amount of agglomerates is present.

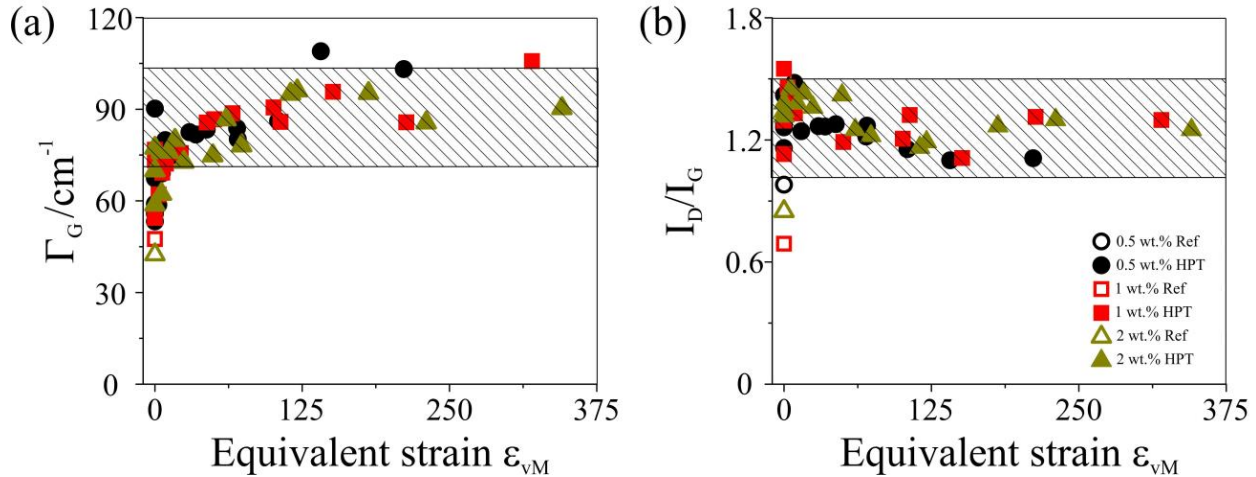

*Figure 4 – (a) Evolution of $\Gamma_G$ as a function of the equivalent strain for different processing conditions. (b) Variation of the $I_D/I_G$ ratio with increasing equivalent strain.*

This is in agreement with the $I_D/I_G$ results (Figure 4b), where the index seems to remain constant in all the cases with the values lying between 1 and 1.5. Again, this quasi steady state is reached at quite low accumulated equivalent strain and remains, to some extent, within the same magnitude. That means that the higher amount of damage is produced at early stages. Since the data do not change significantly after the initial stages of deformation, an extra parameter must be included in the analysis to differentiate the structural disorder. In this case, the examination of the evolution of the G band position would provide a better approach to the identification of the potential amorphization of the $sp^2$ structure. Ferrari and Robertson [32] proposed a straightforward method to evaluate the damage in carbonaceous structures by observing the up and downshifting of the G band. Carbon materials undergo an amorphization trajectory depending on the clustering of the $sp^2$ phase, the bond disorder, the presence of $sp^2$ rings or chains and the $sp^2/sp^3$ ratio [32]. These changes can be tracked and analyzed from the variations in the Raman spectra (i.e. relative intensities, broadening and position shifts of the main peaks). The model consists of three stages, describing the transition from a graphitic to a predominantly $sp^3$ hybridized material. In the first stage, the main structural change is the alteration from a monocrystalline to a polycrystalline phase. Characteristically, the G band presents an upshift to larger wavenumbers (up to 1600 cm$^{-1}$). This upshift is actually a convolution between the original G band (approx. 1580 cm$^{-1}$) and the D' band (approx. 1620 cm$^{-1}$). The second stage comprises a decrease of the G band position towards values around 1510 cm$^{-1}$

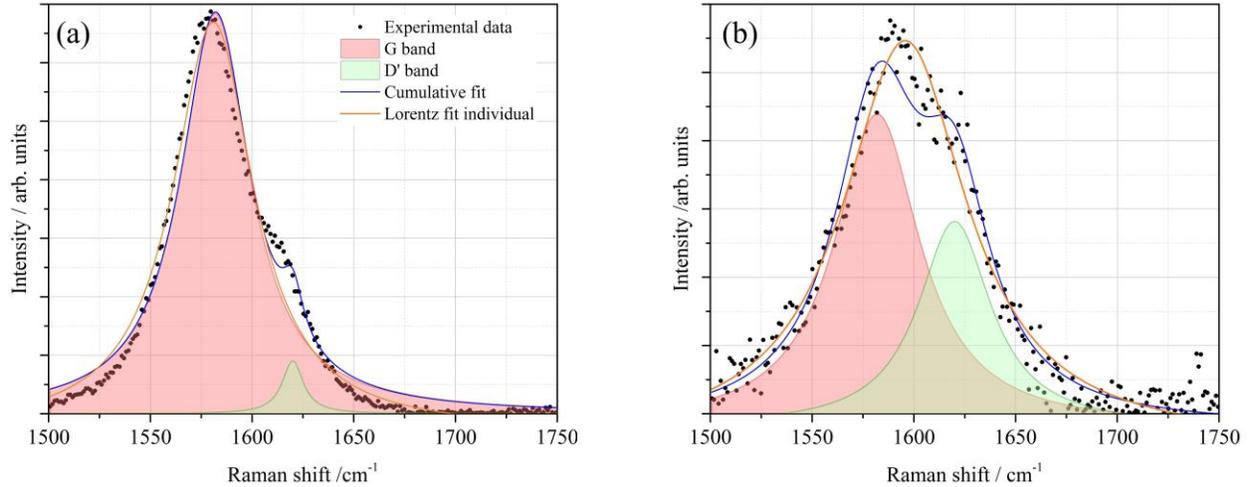

[1], and is related to the accumulation of introduced defects in the graphitic lattice, deriving in a softening of the phonon modes.

Figure 5 shows exemplarily the deconvolution of the upshifted G band in the pristine CNT and in the case of CNT in a highly deformed sample. This peak is a convolution of two, the G aband at 1580 cm$^{-1}$ and the D' band at 1600 cm$^{-1}$ [32].

*Figure 5 – Deconvolution of G peak of (a) Pristine CNT and (b) Ni/0.5 wt.%CNT deformed to an equivalent strain of 195.*

Figure 6 shows the position of the G band for the different samples as a function of the deformation degree. The low CNT-containing samples (0.5 wt.% CNT) show a clear upshifting trend for low deformations, whereas at higher deformation, an inflection point is identifiable. This turning point is located between a von Mises strain of 65 and 130 for 10 turns and between 125 and 250 for 20 turns. In the case of the higher CNT concentrations, the transition occurs at earlier stages of deformation. Specifically, the 1 wt.% samples go through the transition at von Mises strains as low as 50, whereas the 2 wt.% samples show the inflection point at an $\varepsilon_{vM} = 25$. This is evidently related to the amount of available CNT that are subjected to the strain and their distribution. In the case of lower concentrations, the matrix predominantly absorbs the deformation energy, as opposed to the case in which a broad distribution of CNT renders them more vulnerable to the strain.

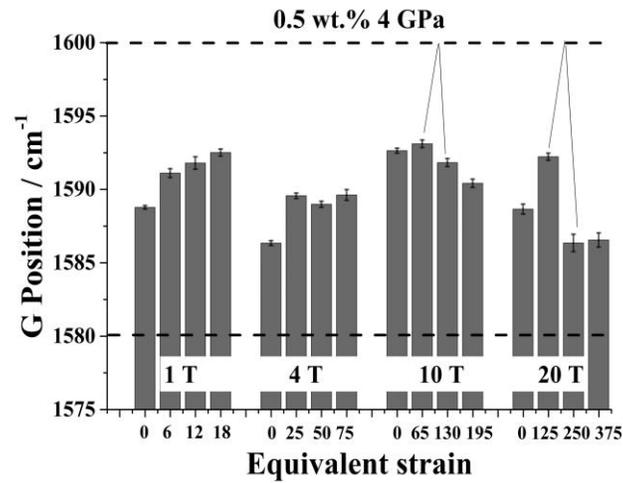

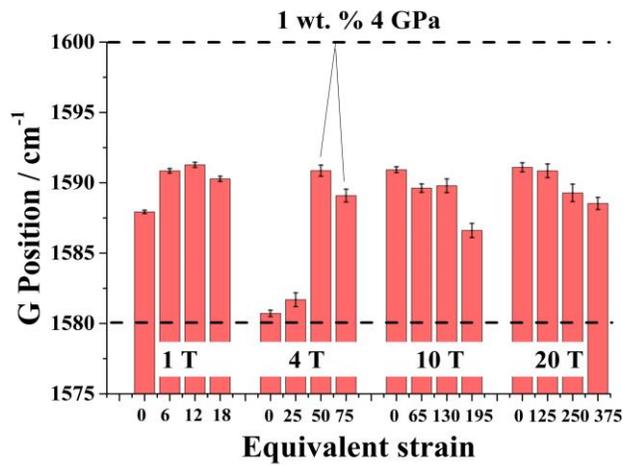

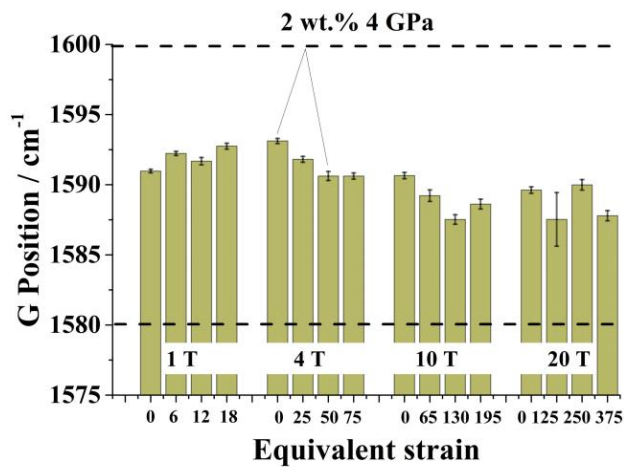

*Figure 6 - Variation of G peak position as a function of the equivalent von Mises strain for the different evaluated CNT concentrations. The dashed lines indicate the lower and upper bounds of the first stage in Ferrari's model.*

In spite of the fact that the CNT undergo an amorphization trajectory as indicated in Figure 6, the G band does not shift to values lower than 1580 cm$^{-1}$, which, according to the aforementioned model proposed by Ferrari et al. and, in the worst case, the nanotubes are at the late periods of the model's first stage.

Concerning the damage mechanisms, it is well known that HPT processing acts on the microstructure of the metallic matrix by inducing large amounts of geometrically necessary dislocations (GND). Since the CNT are placed at the grain boundaries and act as non-coherent bodies with the matrix, the likely strengthening mechanisms would be Orowan looping and grain boundary strengthening [12,19,33]. That means that the GND would round up the second phases, without directly interacting with them. Thus, the most likely damage mechanism acting on the CNT is the matrix plastic flow. Since the CNT are subjected to a strong axial pressure and a shearing movement, it is highly likely that after a certain deformation, the weak structural points of the CNT (structural defects) no longer resist the induced strain and break. The consecution of deformation and breaking of the nanotubes activates the aforementioned amorphization trajectory, sequentially damaging the structure as schematically shown in Figure 7. Furthermore, shorter CNT imply a reduction in the effective load transfer area, which can be detrimental to the mechanical properties of the composites.

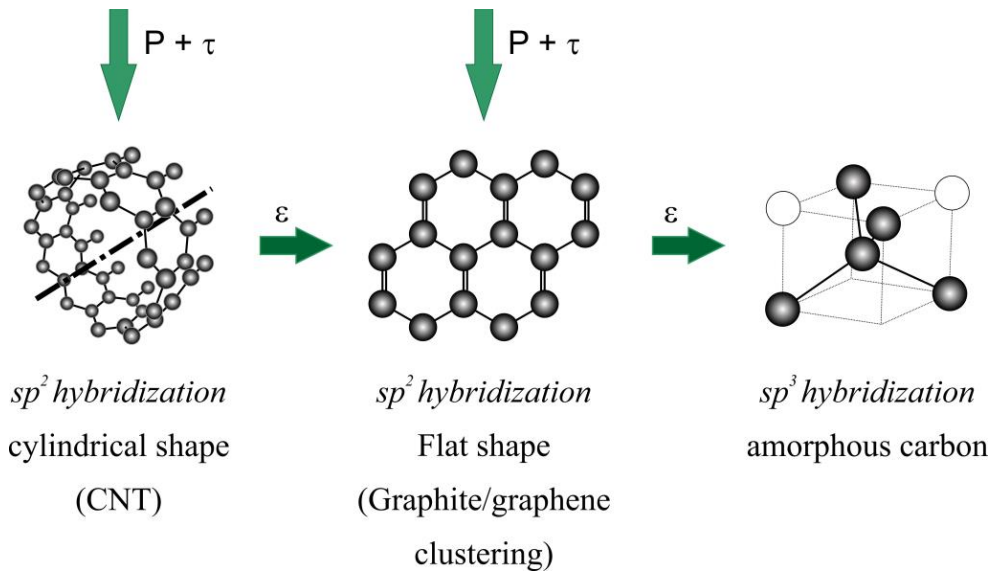

*Figure 7 – Representation of the different stages of the amorphization trajectory that sp$^2$ carbons can undergo.*

Summarizing, after a thorough analysis aided by Raman spectroscopy, it is evident that the carbon nanotubes are strongly affected by the applied strain during deformation, independent of their relative amount in a composite material. It is then evident that an analysis based solely on the well-known defect index ($I_D/I_G$) is insufficient and, for a proper identification of the processing limits, a complementary analysis considering the G band peak width and position is of great importance.

Finally, it has been shown that the processing at room temperature of this type of composites by HPT is possible to a certain extent (defined by the degradation of the nanotubes). The distribution of the agglomerates is improved and a reduction in the mean agglomerate size is achieved. This reduction might play a very important role in the thermal stabilization of the microstructure and subsequently, in the mechanical performance of the composites.

## 4. Conclusions

We have shown that, even though HPT is a useful technique when it comes to the improvement of the distribution of CNT agglomerates in metal matrix composites, it induces irreversible damage on the CNT. In this study, we tracked the defects on the CNT from the *as-sintered* to the *as-deformed* condition for different CNT concentrations and different degrees of deformation. We concluded that the apparent maximum damage, as shown by the evolution of the different Raman parameters here studied and the shifting of the G band, is reached already at early stages of the deformation process, and occurs before the saturation in the microstructural refinement is reached. Specifically, for high CNT concentrations it can start even at a von Mises strain of 25. Furthermore, we were able to correlate the damage mechanism to a well-established carbon amorphization trajectory proposed in the literature. Finally, this study demonstrates that, in order to reliably characterize and track the mechanically induced damage in nanocarbons, the widely used $I_D/I_G$ ratio is not sufficient.

## Acknowledgements

K. Aristizabal wishes to thank the German Academic Exchange Service (DAAD) for their financial support. S. Suarez and K. Aristizabal gratefully acknowledge the financial support from DFG (Grant: SU911/1-1). A. Katzensteiner and A. Bachmaier gratefully acknowledge the financial support by the Austrian Science Fund (FWF): I2294-N36.

The authors want to acknowledge Prof. Volker Presser for providing access to the Raman spectrometer for the measurements.